# Maximally Random Jamming of Two-Dimensional One-Component and Binary Hard Disc Fluids

Xinliang Xu and Stuart A. Rice

Department of Chemistry and The James Franck Institute,
The University of Chicago, Chicago, IL 60637

**Abstract**

We report calculations of the density of maximally random jamming (aka random close packing) of one-component and binary hard disc fluids. The theoretical structure used provides a common framework for description of the hard disc liquid to hexatic, the liquid to hexagonal crystal and the liquid-to-maximally random jammed state transitions. Our analysis is based on locating a particular bifurcation of the solutions of the integral equation for the inhomogeneous single particle density at the transition between different spatial structures. The bifurcation of solutions we study is initiated from the dense metastable fluid, and we associate it with the limit of stability of the fluid, which we identify with the transition from the metastable fluid to a maximally random jammed state. For the one-component hard disc fluid the predicted packing fraction at which the metastable fluid to maximally random jammed state transition occurs is 0.84, in excellent agreement with the experimental value 0.84 ± 0.02. The corresponding analysis of the limit of stability of a binary hard disc fluid with specified disc diameter ratio and disc composition requires extra approximations in the representations of the direct correlation function, the equation of state, and the number of order parameters accounted for. Keeping only the order parameter identified with the largest peak in the structure factor of the highest density regular lattice with the same disc diameter ratio and disc composition as the binary fluid, the predicted density of maximally random jamming is found to be 0.84 to 0.87, depending on the equation of state used, and very weakly dependent on the ratio of disc diameters and the fluid composition, in agreement with both experimental data and computer simulation data.

# I. Introduction

Interest in the character of the so-called random close-packed state [1-3] of matter has grown with the recognition that it is relevant to the understanding of phenomena and systems as diverse as jamming of granular particles, the glass transition, and porous media [4-9]. A random close-packed state of an assembly of spheres is pictured as a collection of particles packed, without ordering, into the densest possible configuration. Given that the spheres have non-zero volume, the highest density configuration cannot have a truly random distribution of particle positions. In fact, when the spheres are jammed together in a fashion that prevents movement, there is short-range order and quasi-long range pair correlations, but not long-range order. For that reason, following Torquato et al [10, 11], we prefer to refer to this particle configuration as the maximally random jammed (MRJ) state. There are several definitions that prescribe quantitative realizations of the qualitative picture of the MRJ state presented above. In this paper we focus attention on the two-dimensional (2D) hard disc fluid, and we identify the transition to the MRJ state with the termination of the metastable branch of the fluid, that is with the limit of stability of the metastable fluid [12]. The theoretical structure we use provides a common framework for description of the hard disc liquid to hexatic, the liquid to hexagonal crystal and the liquid to MRJ state transitions. We will discuss the transition to the MRJ state for both one-component and binary hard disc fluids.

Consider first the one-component hard disc fluid. The best available information from computer simulations and theoretical analyses indicates that as its density is increased the hard disc fluid freezes in two steps, first to a hexatic phase and then to a hexagonal crystal [13]. A theory of the transition to the maximally random jammed state should also be capable of accounting for the fluid to hexatic and hexatic to crystal transitions and be capable of bypassing those transitions so as to describe the metastable fluid state. We previously described a theory of the 2D hard disc fluid to hexatic [14] and hexatic to crystal [15] transitions based on analysis of the nonlinear integral equation describing the inhomogeneous density distribution at phase equilibrium. That analysis takes the form of a search for bifurcation points at which the uniform density of the fluid becomes unstable relative to the density distributions characteristic of the hexatic and/or hexagonal crystal phases. For a specified representation of the structure of the ordered state, the fluid to ordered state phase transition is identified with the lowest density bifurcation point. This identification picks out the transition between equilibrium states of the fluid and solid. The hard disc fluid densities at which these transitions occur, expressed as packing fractions $\eta = N\pi\sigma^2/4A$, are found to be $\eta^{Hexatic} = 0.691$ and $\eta^{Xtal} = 0.708$, respectively, both in very good agreement with the results obtained from simulations reported by Mak [13], namely 0.703 and 0.723, respectively. We show in this paper that if the fluid solution for the density is followed past the freezing transition into the region in which the fluid is unstable relative to the crystal another bifurcation point is found, which we identify with the transition to the MRJ state. We find, for the one-component hard disc fluid, $\eta^{MRJ} = 0.84$, in quantitative agreement with the experimental value, $0.84 \pm 0.02$ [16]. The packing fraction of the 2D close packed hexagonal lattice is 0.9069.

Consider now the binary hard disc fluid and the binary hard disc solid. There are many binary hard disc ordered crystals, each characterized by a specific ratio of disc diameters and disc mole fractions. We restrict attention to the case that the smaller hard disc cannot fit in the interior space of three tangent large discs, which is true for $\alpha \equiv \sigma_2/\sigma_1 \geq 0.155$. A few examples of such lattices are displayed in Fig. 1. For the structures displayed in Fig. 1, $\alpha$ ranges from 0.637 (upper left) to 0.217 (lower right) and the corresponding mole fraction of small discs ranges from 0.500 to 0.800, yet the packing fractions only vary from 0.9110 to 0.9331. When the binary hard disc fluid has the same disc diameter ratio and composition as an ordered solid we expect freezing to that solid to occur; if the composition of the fluid deviates from that of the ordered solid, we expect freezing to generate two or more ordered solid phases with different compositions, one of which may be pure component 1 or 2. The observation we wish to emphasize is that the densities of binary hard disc crystals are so little different that it is reasonable to anticipate that the densities of the corresponding maximally jammed states of the system will have little variation with disc diameter ratio and mole fraction. Indeed, Bideau and coworkers [17] have reported the results of mechanical simulations of binary hard disc fluids for $0.8 \geq \alpha \geq 0.2$, and mole fraction of the larger disc $0.9 \geq x_1 \geq 0.1$. They report that the transition to the MRJ state occurs at $\eta^{MRJ}_{MIX} = 0.84 \pm 0.02$ independent of fluid composition and of ratio of disc diameters. Barker and Grimson [18] have reported the results of extensive computer simulations of the transition to the MRJ state in binary hard disc fluids. Their results also show that this transition occurs at a density that is sensibly independent of fluid composition and ratio of disc diameters, with $\eta^{MRJ} = 0.818$ for the one-component fluid and with average value $\eta^{MRJ}_{MIX} = 0.815$ for the composition range $0.01 \leq x_1 \leq 1$. Although we have used the words "sensibly independent", the simulation data are sufficiently precise to show a very weak dependence of $\eta^{MRJ}_{MIX}$ on fluid composition and ratio of disc diameters. We regard the observation that $\eta^{MRJ}_{MIX}$ is nearly independent of fluid composition and ratio of disc diameters more remarkable than the observation of a very weak variation of $\eta^{MRJ}_{MIX}$ with these parameters. Finally, we note that the simulation calculations of $\eta^{MRJ}_{MIX}$ show a very weak dependence on the shape of the simulation box and the number of particles, suggesting that the difference between the simulated and experimental transition densities may not be significant.

## II. Theoretical Analysis: General Remarks

The starting point for our analysis is an exact equation for the singlet density distribution, $\rho(1)$, as a function of position in space. This equation, derived independently by Arinshtein [19] and by Stillinger and Buff [20], has the form

$$\ln \frac{\rho(1)}{z} = \sum_{k=1}^{\infty} \frac{1}{k!} \int S_{k+1}(1,..,k+1) \prod_{i=2}^{k+1} \rho(i) d(i) \qquad (2.1)$$

where $S_{k+1}(1,..,k+1)$ is the sum of all irreducible Mayer diagrams of order $k + 1$, $z$ is the fugacity of the system and $d(i)$ denotes integration over the coordinates of particle $i$. The right hand side of Eq. (2.1) is the generating functional for the set of $n$-particle direct correlation functions:

$$c_n(1,...n) = \frac{\delta^{n-1} \ln(\rho(1)/z)}{\delta \rho(2)...\delta \rho(n)} \qquad (2.2)$$

Then, defining the generalized free energy functional $F(1; \rho(1)) \equiv \ln(\rho(1)/z)$ we can rewrite (2.2) in the form of a functional Taylor series relating the position dependent densities in two phases that differ in mean density by $\Delta\rho$. When one of the phases is selected to be the liquid and the other is crystalline with reciprocal lattice vector set $\{\mathbf{G}\}$ and unit cell volume $\Delta_S$ Eq. (2.1) assumes the form

$$F(\rho_L, \Delta\rho(\mathbf{r}_1)) = F(\rho_L, 0) + \sum_{n=1}^{\infty} \frac{1}{n!} \int c_n^L(1,...,n+1) \prod_{i=2}^{n+1} \rho(i) d(i), \qquad (2.3)$$

$$\rho_L + \Delta\rho(\mathbf{r}) = \rho_L \left( 1 + \phi_0 + \sum_{\mathbf{G}} \phi_{\mathbf{G}} \exp(i\mathbf{G} \bullet \mathbf{r}) \right), \qquad (2.4)$$

in which $c_n^L(1,...,n+1)$ is the *n*-particle direct correlation function of the liquid. The coefficients $\{\phi_\mathbf{G}\}$,

$$\phi_\mathbf{G} = \frac{1}{\Delta_S} \int d\mathbf{r} \frac{\Delta\rho(\mathbf{r})}{\rho_L} \exp(-i\mathbf{G} \bullet \mathbf{r}), \qquad (2.5)$$

serve as the order parameters for the transition.

Our analysis of the transition from the metastable fluid to the MRJ state draws on the results of a very detailed study, by Mayer, of the general properties of integral equations for equilibrium distribution functions, and on an interpretation of the behavior of the metastable fluid [21]. Mayer derived a number of exact relationships between distribution functions at different fugacities, say $z_\alpha$ and $z_\beta$, and an interpretation of the solutions of the integral equations for the distribution functions under the conditions of phase equilibrium. In particular he showed that Eq. (2.1) and its generalization to non-fluid systems possess unique solutions in the one-phase regions supported by the system, and that the equations have solutions for all values of *z* except those, $z_\gamma$, at which phase changes occur. The equations do not describe the two-phase region since specification of the fugacity of a system does not specify the amounts of the two phases in equilibrium. The unique values of $z_\gamma$ for which phase transitions occur are obtained from the eigenvalues of an equation involving a kernel that is related to the correlation function. In general, the distribution functions of the system are different in different phases, and do not approach one another as $z_\alpha$ or $z_\beta$ approaches $z_\gamma$, the fugacity at the phase transition; when $z_\alpha = z_\gamma$ or $z_\beta = z_\gamma$ the solutions to the integral equation change character. The location of the phase transition can, therefore, be determined by finding where the solution to the non-linear integral equation (2.1) for the distribution function, or a surrogate derived from it, changes character, i.e., bifurcates with a discontinuity in the density. Of course, the accuracy of this procedure is compromised by any approximations that reduce the accuracy of Eq. (2.1) or Eq. (2.3).

The interpretation of the behavior of the metastable fluid we use is based on the nonlinear integral equation obtained by truncation of the right hand side of Eq. (2.3) at the level of the pair direct correlation function. This nonlinear integral equation has a number of properties, one of which is pertinent to the subject of this paper. Bagchi, Cerjan and Rice [22] showed that if the fluid solution for the density is followed past the bifurcation point associated with the liquid-to-crystal transition, into the region in which the fluid is unstable relative to the crystal, another bifurcation point is found. If the structure of the crystal at the density of the second bifurcation point is such that it generates maximum covering of the space, that bifurcation point locates the limit of stability of the metastable fluid. The density of the metastable fluid at that bifurcation point is less than the density of the crystal that maximally covers the space, which leads to the identification of that bifurcation point with the transition to the MRJ state. This interpretation is supported by the exact solution of the nonlinear integral equation for the one-dimensional hard rod fluid, and by the numerical location of the bifurcation point for the three-dimensional hard sphere fluid. In the former case the density at the bifurcation point corresponds to close packing of the hard rods, in agreement with the absence of a phase transition in a one-dimensional system of particles with hard-core interaction and in agreement with identification of the limit of stability of the liquid with a divergent elastic modulus. In the latter case the theory predicts the volume fraction for the transition to the MRJ state in excellent agreement with simulation data, namely $\phi^{MRJ} = N\pi\sigma^3/6V = 0.63$. As will be shown below that theory with some modern extensions, when applied to the one-component hard disc fluid, leads to the prediction $\eta^{MRJ} = 0.84$, in quantitative agreement with the experimental value, $0.84 \pm 0.02$.

### III. Bifurcation Analysis: One-Component Hard Disc Fluid

To find the bifurcation of solutions to Eq. (2.3), which is exact, we truncate the right hand side at the level of the pair direct correlation function. This approximation leads to an error of order $\phi_0$, the density difference between the phases at the transition point. We later show that for the hard disc fluid this error is less than 0.5%. With this approximation Eq. (2.3) becomes

$$\rho_L + \Delta\rho_L = \rho_L \exp\left[\int c_2^L(\mathbf{r}_{12})\Delta\rho(\mathbf{r}_2)d\mathbf{r}_2\right], \tag{3.1}$$

noting that $c_2^L(\mathbf{r}_1,\mathbf{r}_2) = c_2(\mathbf{r}_{12})$ in the homogeneous liquid and $z_L = z_S$ at the transition point. Combination of Eq. (2.4) and (3.1) then leads to

$$1+\phi_0 = \frac{1}{\Delta_S}\int_{\Delta_S} d\mathbf{r}_1 \exp(\sigma_0)\exp\left(\sum_{\mathbf{G}}\sigma_{\mathbf{G}}\psi_{\mathbf{G}}\xi_{\mathbf{G}}(\mathbf{r}_1)\right), \tag{3.2}$$

$$\psi_{\mathbf{G}_n} = \frac{\int_{\Delta_S} d\mathbf{r}_1 \xi_{\mathbf{G}_n}(\mathbf{r}_1)\exp\left(\sum_{\mathbf{G}}\sigma_{\mathbf{G}}\psi_{\mathbf{G}}\xi_{\mathbf{G}}(\mathbf{r}_1)\right)}{\int_{\Delta_S} d\mathbf{r}_1 \exp\left(\sum_{\mathbf{G}}\sigma_{\mathbf{G}}\psi_{\mathbf{G}}\xi_{\mathbf{G}}(\mathbf{r}_1)\right)}, \tag{3.3}$$

$$\psi_{\mathbf{G}} = \frac{\phi_{\mathbf{G}}}{1+\phi_{\mathbf{G}}},$$

$$\xi_{\mathbf{G}}(\mathbf{r}) = \exp(i\mathbf{G}\bullet\mathbf{r}),$$

$$\sigma_0 = \rho_L\phi_0\int d\mathbf{r} c^L(\mathbf{r}) = \phi_0\left(1-\frac{1}{S(0)}\right), \tag{3.4}$$

$$\sigma_{\mathbf{G}} = \rho_L(1+\phi_0)\int d\mathbf{r}\xi_{\mathbf{G}}(\mathbf{r})c^L(\mathbf{r}) = (1+\phi_0)\left(1-\frac{1}{S(\mathbf{G})}\right)$$

with $\mathbf{G}_n$ the $n$ th reciprocal lattice vector, and $S(0)$ and $S(\mathbf{G})$ the structure functions of the hard disc system evaluated at zero and at reciprocal lattice vector $\mathbf{G}$.

A visualization of the conditions for the solution of Eq. (3.3) is presented in Fig. 2 for the simplest case in which only the term corresponding to the first reciprocal lattice vector is retained in Eq. (2.4). The bifurcation condition is then

$$\sigma_{\mathbf{G}}(\rho_L, \phi_0) = (1+\phi_0)\left(1 - \frac{1}{S(\mathbf{G})}\right) = \sigma_{\mathbf{G}}^*. \tag{3.5}$$

Eq. (3.5) locates $\sigma_{\mathbf{G}}^*$ at the point of tangency of Eq. (3.3) and the straight line with slope $1/\sigma_{\mathbf{G}}$. When higher accuracy is sought, by inclusion of higher order reciprocal lattice vectors, one must solve Eq. (3.3) simultaneously with a set of equations, one for each of the reciprocal lattice vectors included. These equations are

$$\sigma_{\mathbf{G}_n}(\rho_L, \phi_0) = (1+\phi_0)\left(1 - \frac{1}{S(\mathbf{G}_n)}\right) = \sigma_{\mathbf{G}_n}^*, \; n = 1, 2, \ldots \tag{3.6}$$

As just described, this procedure and selection of the lowest density bifurcation point defines the liquid-to-crystal, or liquid-to-hexatic transition. And, as mentioned in the Introduction, when applied to the hard disc fluid, with suitable representations of the order in the hexatic phase and the order in the hexagonal crystal phase, the predicted liquid densities at the transition points, $\eta^{Hexatic} = 0.691$ and $\eta^{Xtal} = 0.708$ are in very good agreement with available simulation data [13], namely $\eta^{Hexatic} = 0.703$ and $\eta^{Xtal} = 0.723$.

The transition to the MRJ state is different from the liquid-to-hexatic and liquid-to-crystal transitions in that it proceeds from the metastable fluid at a density greater than that at the either the fluid-to-hexatic or the fluid-to-crystal transitions. The fluid-to-crystal transition generates an ordered solid with density less than that at close packing of the particles. Although the fluid-to-crystal transition can be located for any particular crystal lattice, each lattice with different stability relative to the fluid and other crystal lattices, our analysis of the transition to the MRJ state requires that the particular fluid-to-crystal transition that is bypassed involve that crystal structure that maximally covers the space when the particles are in contact. For the 2D hard disc system this is the hexagonal close-packed lattice.

A schematic illustration of the consequences of following the metastable fluid branch of the nonlinear integral equation is displayed in Fig. 3. Figure 3a compares the free energy, as described by the Landau theory of first order phase transitions, as a function of an order parameter $q$. The blue line in this figure represents the free energy at a first order transition point, with $F(0) = F(q^*)$ and both $q = 0$ and $q = q^*$ minima of the free energy. The red line represents the free energy at the limit of stability of the fluid; the derivative of the free energy with respect to $q$ is zero at $q = 0$ and that point is not a local minimum of the free energy. Figure 3b displays a sketch of Eq. (3.3) for the same conditions; the slope of the blue line gives the bifurcation condition for the liquid-to-crystal transition and the slope of the red line, which is equal to 1, gives the condition for the limit of stability. The conditions that define the limit of stability are Eq. (3.2) and

$$\sigma_{\mathbf{G}_1}(\rho_L, \phi_0) = (1 + \phi_0)\left(1 - \frac{1}{S(\mathbf{G}_1)}\right) = 1,$$
$$\sigma_{\mathbf{G}_n}(\rho_L, \phi_0) = (1 + \phi_0)\left(1 - \frac{1}{S(\mathbf{G}_n)}\right) = \sigma^*_{\mathbf{G}_n} \text{ for } n \geq 2.$$
(3.7)

We have generated numerical solutions to Eqs. (3.2) and (3.7) for several choices of sets of reciprocal lattice vectors, using the Baus-Colot representation [23] of the direct correlation function of a two dimensional hard-disc fluid. This representation of the direct correlation function is known to be very accurate in the super-cooled fluid regime. The result of our calculation is the prediction that the transition to the MRJ state occurs at a packing fraction

$$\eta^{MRJ} = 0.84, \quad \phi_0 = 0.003, \quad S(\mathbf{G}) \approx 300,$$
(3.8)

in excellent agreement with available experimental data [17], $\eta^{MRJ} = 0.84 \pm 0.02$. The calculated value of $\eta^{MRJ}$ and the value of $S(\mathbf{G})$ at the first peak of the structure function are found to be insensitive to the number of order parameters included in the calculation (see Table 1). We also note that our calculations show that there is negligible change in density at the transition to the MRJ state, justifying our truncation of Eq. (2.3).

Table 1. Dependence of Fluid-to-Maximally-Random Jammed State Transition Density on Number of Order Parameters

|  | $\eta^{MRJ}$ | $S(\mathbf{G})$ |
|---|---|---|
| 1 order parameter | 0.8407 | ~300 |
| 3 order parameters | 0.8407 | ~300 |
| 5 order parameters | 0.8408 | ~300 |

## IV. Theoretical Analysis: Binary Hard Disc Fluid

Application of our analysis to the binary hard disc fluid requires information concerning the structure factor of the fluid and, for specified diameter ratio of the discs and composition of the binary fluid, the structure of the ordered phase with greatest coverage of the plane. The latter information is required because our analysis of the transition to the MRJ state implies that the particular fluid-to-crystal transition that is bypassed involves that crystal structure that maximally covers the space. Our analysis of the transition to the MRJ state in a binary hard disc fluid is more approximate than for the pure hard disc fluid because of uncertainties in the fluid direct correlation function and because we are able to carry out the calculations only to the one order parameter level, corresponding to locating the largest amplitude peak of the structure function of the binary crystal structure that maximally covers the space.

We restrict attention to binary hard disc mixtures that have disc diameter ratio and disc mole fractions corresponding to an ordered binary hard disc crystal. Then the extension of Eq. (2.4) is the set of equations ($i = 1,2$)

$$\rho_i(\mathbf{r}) = \rho_i^L + \Delta\rho_i^L(\mathbf{r}) = \rho_i^L\left(1 + \phi_{i0} + \sum_{\mathbf{G}} \phi_{i\mathbf{G}} \exp(i\mathbf{G} \cdot \mathbf{r})\right), \quad (4.1)$$

and Eq. (2.2) is generalized to

$$c_{ij}(\mathbf{r}_1, \mathbf{r}_2) = \frac{\delta \ln(\rho_i(\mathbf{r}_1)/z)}{\delta \rho_j(\mathbf{r}_2)}, \quad (4.2)$$

and Eq. (3.1) becomes the set of equations ($i = 1,2$)

$$\rho_i^L(\mathbf{r}_1) + \Delta\rho_i^L(\mathbf{r}_1) = \rho_i^L \exp\left[\sum_{j=1,2} \int c_{ij}(\mathbf{r}_1, \mathbf{r}_2) \Delta\rho_i^L(\mathbf{r}_2) d\mathbf{r}_2\right] \quad (4.3)$$

The liquid-to-crystal phase transition of the one component liquid is identified with the bifurcation point of Eq. (3.5) at which the density distribution changes from a constant to a periodic function. Similarly, the liquid-to-crystal transition of a binary mixture is identified with the density at which simultaneous bifurcations of the two equations in the set (4.3) occur. When only the term corresponding to the first reciprocal lattice vector is retained in equation (4.1) this condition is met for the binary mixture when $\lambda_\sigma(\mathbf{G}) = 1$, where $\lambda_\sigma(\mathbf{G})$ is the largest eigenvalue of the matrix

$$\begin{pmatrix} \sigma_{11\mathbf{G}} & \sigma_{12\mathbf{G}} \\ \sigma_{21\mathbf{G}} & \sigma_{22\mathbf{G}} \end{pmatrix}; \quad \sigma_{ij\mathbf{G}} = \rho_i^L(1+\phi_{i0})c_{ij}(\mathbf{G}); \quad c_{ij}(\mathbf{k}) = \int d\mathbf{r} \exp(i\mathbf{k}\bullet\mathbf{r})c_{ij}(\mathbf{r}). \qquad (4.4)$$

Our analysis of the transition to the MRJ state in the one component hard disc fluid revealed that the location of the bifurcation point is primarily determined by the dominant peak in the structure factor $S(\mathbf{k})$, as shown (Table 1) by the insensitivity of $\eta^{MRJ}$ and the peak value of $S(\mathbf{k})$ to the number of order parameters included in the calculation. We assume the same insensitivity will be characteristic of the transition to the MRJ state of a binary mixture. To carry out the calculation we introduce the matrix of structure factors defined by

$$[\![S(\mathbf{k})]\!] = \begin{pmatrix} S_{11}(\mathbf{k}) & S_{12}(\mathbf{k}) \\ S_{21}(\mathbf{k}) & S_{22}(\mathbf{k}) \end{pmatrix}; \quad S_{ij}(\mathbf{k}) = \left[\mathbf{I} - \widetilde{\mathbf{C}}(\mathbf{k})\right]^{-1}_{ij}; \quad \left[\widetilde{\mathbf{C}}(\mathbf{k})\right]_{ij} = \left(\rho_i \rho_j\right)^{1/2} \widetilde{c}_{ij}(\mathbf{k}). \quad (4.5)$$

It is the largest eigenvalue of $[\![S(\mathbf{k})]\!]$, denoted $\lambda_S(\mathbf{k})$, that determines the location of the bifurcation to the MRJ state. Indeed, it is sufficiently accurate to assume that $\lambda_S(\mathbf{k}) \colon 300$ because at the packing fraction corresponding to this value of $\lambda_S(\mathbf{k})$ a 0.1% increase of the packing fraction corresponds to $\lambda_S(\mathbf{k})$ with twice this value.

To obtain the matrix of structure factors of a given binary mixture of hard discs we use a procedure proposed by Barrat, Xu, Hansen and Baus [24]. This procedure is a generalization to binary mixtures of the rescaling protocol of the Baus-Colot representation of the pair direct correlation function of a one-component hard disc fluid. The protocol has two decoupled steps: (1) a guess of the form of the direct pair correlation function, and (2) a guess of the form of the equation of state. Consider, for example, the one-component hard disc fluid. The direct correlation function is first written in the form $c(r) = c(r=0)f(r/a(\eta))$ in which $c(r=0)$ sets the overall scale of the function and $f(r/a(\eta))$ is a rescaling of the exact low-density expression for the direct correlation function with scaling factor $a(\eta)$ a function of the packing fraction. With the additional assumption that $\int c(r)g_2(r)dr \approx 0$, both $c(r)$ and $f(r/a(\eta))$ can be calculated from the equation of state. We note that this assumption is exact in the Percus-Yevick approximation. Baus and Colot then show that using the parameterized equation of state

$$\frac{p}{\rho k_B T} = \frac{1 + c_1\eta + c_2\eta^2}{(1-\eta)^2}, \qquad (4.6)$$

(with $c_1$ and $c_2$ calculated to yield the correct second and third virial coefficients) to calculate $a(\eta)$, yields very good predictions of the structure and the liquid-to-crystal transition density. As shown in Section III, it also yields a very good prediction of the density at which the transition to the MRJ state occurs.

We have used the same rescaling procedure with the direct correlation functions proposed by Barrat et al for the binary hard disc mixture. However, representations of the equation of state of the binary hard disc mixture are less accurate than those for the one-component hard disc fluid. It is known that when extended to the binary hard disc mixture the analogue Eq. (4.6) overestimates the pressure at small packing fraction and underestimates the pressure close to the MRJ state [25, 26]. Several different forms of equation of state have been proposed to reduce these discrepancies [27, 28]. We propose the following approximate equation of state for the binary hard disc mixture, expected to be accurate in the regime close to the MRJ density:

$$\frac{p}{\rho k_B T} = \frac{1 + c_1\eta + c_2\eta^2}{\left(1 - \eta\langle\sigma^2\rangle/\langle\sigma\rangle^2\right)}. \qquad (4.7)$$

In Eq. (4.7), $\langle\sigma\rangle = x_1\sigma_1 + x_2\sigma_2$ and $\langle\sigma^2\rangle = x_1\sigma_1^2 + x_2\sigma_2^2$. Using Eq. (4.6) and Eq.(4.7) we have generated numerical values of the matrix elements $S_{ij}(\mathbf{k})$ for binary mixtures with $x_1 = x_2 = 0.5$ and various disc diameter ratios. And by replacing $S(\mathbf{G})$ in Eq. (3.4) with $\lambda_S(\mathbf{G})$, the largest eigenvalue of the matrix $[\![S(\mathbf{k})]\!]$ at $\mathbf{k} = \mathbf{G}$, we are able to predict, at the one order parameter theory level, the value of $\eta^{MRJ}$ for a binary hard disc mixture with a specified disc diameter ratio. The value found for the mixture with $\alpha = 0.8$ and $x_1 = x_2 = 0.5$, $\eta^{MRJ} = 0.84$, is in excellent agreement with the experimental value obtained by Bideau et al [17]. The superiority of predictions based on Eq. (4.7) to those obtained with several other choices of equation of state is demonstrated by the entries in Table 2.

Table 2. Predicted values of $\eta^{MRJ}$ obtained with various equations of state, with the direct correlation function obtained by the procedure of Barrat et al, except that developed by Rosenfeld, which is a linear interpolation of the analytical solutions to the equation of state under the PY approximation in 1 dimension and 3 dimensions. All of the equations of state, except that developed by Rosenfeld, predict $\eta^{MRJ} = 0.841$ when α = 1. However, only the equation of state represented in Eq. (4.7) provide an accurate prediction of $\eta^{MRJ}$ when α = 0.8.

| Equation of State | α = 1 | | α = 0.8 | |
|---|---|---|---|---|
| | $\eta^{MRJ}$ | $S(\mathbf{k})$ at MRJ | $\eta^{MRJ}$ | $S(\mathbf{k})$ at MRJ |
| Barrat et al [24] | 0.841 | ~300 | 0.902 | ~300 |
| Henderson [29] | 0.841 | ~300 | 0.893 | ~300 |
| Jenkins-Mancini [27] | 0.841 | ~300 | 0.897 | ~300 |
| Rosenfeld [30] | 0.754 | ~300 | 0.789 | ~300 |
| Santos-Yuste-Lopez [28] | 0.841 | ~300 | 0.903 | ~300 |
| Xu-Rice [Eq. (4.7)] | 0.841 | ~300 | 0.842 | ~300 |

We now focus attention on predictions of the MRJ packing fraction, as a function of disc diameter ratio, obtained using Eq. (4.7). Figure 4 shows these predictions of $\eta^{MRJ}$ for a mixture with $x_1 = x_2 = 0.5$ for the range $1 > \alpha > 0.7$. Clearly, $\eta^{MRJ}$ is found to be almost independent of $\alpha$ except for the range $0.72 \leq \alpha \leq 0.76$. A study of the position of the peak of $\lambda_S(\mathbf{k})$ as a function of $\alpha$ (see Fig. 5), and of the structures of possible binary disc crystal lattices as a function of $\alpha$, reveals that the lattice that maximally covers the plane changes in this range of $\alpha$. We infer that the deviation from a nearly constant value of $\eta^{MRJ}$ seen in Fig. 4 is a consequence of competition between these two lattices and the corresponding peaking values of $\lambda_S(\mathbf{k})$ in the range $0.72 \leq \alpha \leq 0.76$, which is not accounted for by the one order parameter theory we have used.

The inference that there is a change in the structure of the binary crystal that maximally covers the plane in this diameter ratio regime is supported by results reported by Fejes Toth [31]. He examined a large number of ordered arrangements of binary disc mixtures with various diameter ratios that he considered to be "good" in the sense that they appear to maximally cover the plane. For the range $1 \geq \alpha \geq 0.645$ all the ordered structure have the packing fraction $\pi/\sqrt{12} = 0.9069$. In the range $0.645 \geq \alpha \geq 0.637$ there is a continuous transition between two ordered lattices that differ only in the percolation of contacts of the large discs, and the packing fraction increases to 0.9110. We suggest that this transition between most densely packed binary lattices is the one responsible for the competing maximum eigenvectors that leads to the deviation seen in Fig. 4 and the discontinuity seen in Fig. 5. The difference between the diameter ratio at which the transition occurs obtained from our one order parameter analysis and that found by Fejes Toth we attribute to the approximations in our theory and the restriction of the bifurcation analysis to the one order parameter level.

There is another approach to approximating the equation of state of the hard disc binary mixture, to be used as above to calculate $\eta^{MRJ}$, that is worth discussing. Stillinger, Torquato and coworkers [10, 32] have shown that the MRJ state exhibits hyperuniform long-range order that is characterized by vanishing infinite-wavelength local volume fraction fluctuations and divergent elastic moduli. The latter feature is captured by the free volume equation of state [10]

$$\frac{p}{\rho k_B T} = \frac{2}{1 - \eta/\eta_{CP}}, \qquad (4.8)$$

where we take $\eta_{CP}$ to be the packing fraction of the close packed lattice as calculated by Fejes Toth [31]. There is numerical evidence (quoted in Ref. 10) that near the MRJ density the free volume model yields an accurate hard disc equation of state. The use of Eq. (4.8) and the procedure described above yields sensible structure factors near $\eta^{MRJ}$ for binary mixtures with different α, and makes plausible predictions for $\eta^{MRJ}$ at all α, albeit with a slightly different value, $\eta^{MRJ} = 0.857$ at α = 1, from that found in Section III.

**V. Discussion**

A feature of the analysis presented in this paper is the inclusion, in a common formalism, of the hard disc liquid-to-hexatic, liquid-to-crystal and metastable liquid to maximally random jammed state transitions. This analysis is consistent with the views presented by Stillinger, Torquato and coworkers [10] concerning the existence of a range of jamming densities rather than a unique jamming density. Our formalism identifies particular bifurcations of the solutions of the integral equation for the inhomogeneous single particle density with particular structural transitions. In each case, having selected the ordered structure against which the fluid stability is tested, it is the bifurcation point at lowest density that is identified with the liquid to hexatic, liquid to crystal, and liquid to maximally random jammed state transitions. The liquid to hexatic and liquid to crystal transitions are between states at equilibrium. That is not the case for the liquid to MRJ state transition. That transition is accessed by exploiting the fact that bypassing the bifurcation point corresponding to the fluid-crystal transition that occurs at lowest density, and thereby ignoring the transition to the distribution of particles that has the lowest free energy, allows the constrained metastable fluid to be described up to the limit of stability of the fluid phase. Since all jammed states lie outside the domain of stability of the fluid phase, our identification of the density of the transition to the MRJ state does not exclude the existence of jammed states with higher density, as shown by Stillinger, Torquato and coworkers.

There are subtle but distinctive differences between the pair correlation functions of the metastable fluid and the MRJ state, a necessary condition for our analysis. As far as short-range order is concerned, typically, the pair correlation function of an amorphous solid, or glass, has a fairly sharp first peak, a split second peak, and third and subsequent peaks that, although broad, are more pronounced than in the parent liquid. Arguably more important, as shown by Stillinger, Torquato and coworkers, the MRJ state exhibits quasi-long range pair correlations and associated divergences of elastic constants [10, 32]. Given that the several equations of state represented in Eqs. (4.6), (4.7) and (4.8) all predict a divergence of the pressure, albeit at different packing fractions, the procedure described in Section IV to determine the direct correlation function of the dense metastable fluid likely builds in, at some level of approximation, the latter feature of the MRJ state. We note that Eq. (3.7) is a condition on the several $S(\mathbf{G}_n)$ that have contributions from all particle separations, and at the transition to the MRJ state the value of $S(\mathbf{G}_1) \approx \phi_0^{-1}$ becomes very large. That large value results from the large separation behavior of the pair correlation function.

Finally, we recall that Kozak, Brzezinski and Rice [33] have examined the conjecture that in a 2D system of hard discs the densities at which the fluid to hexatic and hexatic to crystal transitions occur can be correlated with the packing densities of tessellations (patterned networks) that span the 2D space, and they describe possible tessellations that meet this criterion. They argue that said tessellations do not actually occur in the 2d hard disc fluid, but that the densities at which the fluid to hexatic and hexatic to crystal transitions occur might be signatures of the existence of nearby tessellations that completely span the 2D space, i.e., ghost configurations that parallel the change in character of the solutions to the integral equation for the inhomogeneous density distribution function. Taking the same point of view, we note that Williams [34] has argued that a tessellation of 2D space with rhombuses that have average internal angles of 105º and 75º, uniformly distributed within the ranges 90º-120º and 90º-60º respectively, has a packing density of 0.813. Williams identifies this packing density with the MRJ state. The bifurcation from the metastable fluid to the MRJ state might then be regarded as the signature of the nearby irregular rhombic tessellation that spans the 2D space.

## VI. Acknowledgements

X. L. Xu thanks Dr. X. Cheng for providing unpublished experimental results and for helpful discussions. This work has been supported by the National Science Foundation funded MRSEC Laboratory at The University of Chicago (DMR-0820054).

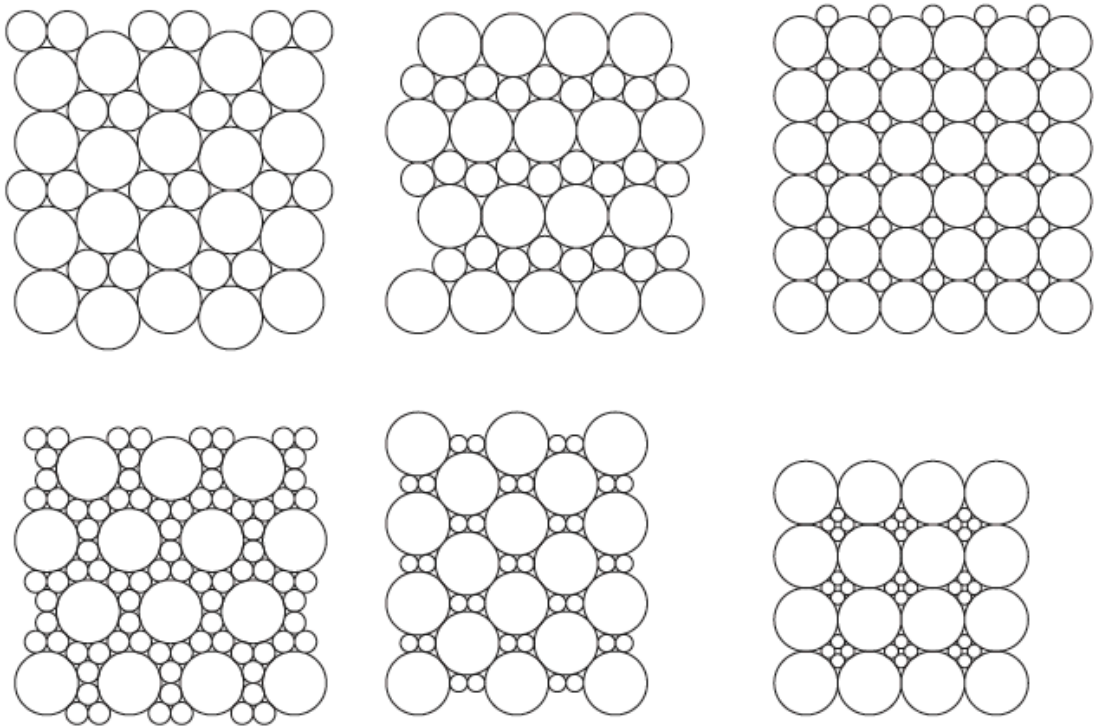

Figure 1. Six ordered binary hard disc lattices (abstracted from Ref. 31). These have the following values of the disc diameter ratio, mole fraction of small discs and packing fractions. From left to right in the top row, then left to right in the lower row: $\alpha$ = 0.6372, 0.5333, 0.4142, 0.3492, 0.2808, 0.2168; $x_2$ = 0.500, 0.667, 0.500, 0.857, 0.667, 0.800; $\eta_{CP}$ = 0.9110, 0.9142, 0.9202, 0.9262, 0.9300, 0.9331.

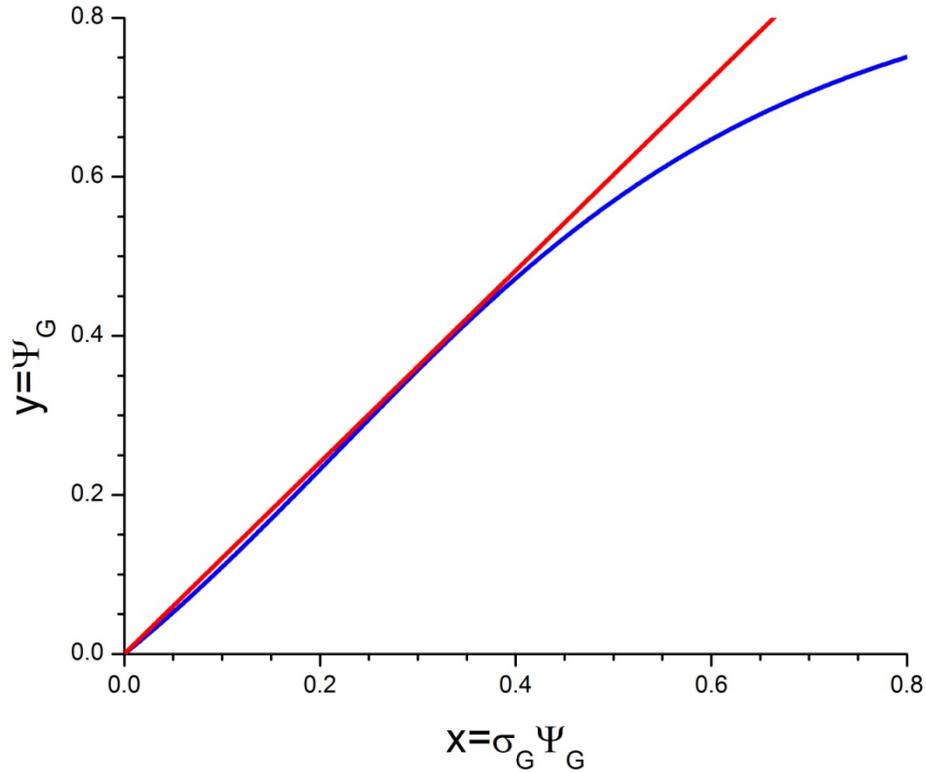

Figure 2. An illustration of the bifurcation condition (see Eq. (3.5). Denoting $x = \sigma_G \psi_G$ and $y = \psi_G$ Eq. (3.3) can be written in the form $y = (x/\sigma_G) = \int \xi_G(\mathbf{r}) \exp(x\xi_G(\mathbf{r})d\mathbf{r} / \int \exp(x\xi_G(\mathbf{r})d\mathbf{r}$. The first term on the right hand side is shown as the red straight line with slope $1/\sigma_G$ and the second term is shown as the blue curve. As the blue curve is fixed, the liquid-crystal phase transition requires a small enough slope, that is, by equation (3.5) large enough $S(\mathbf{G})$, that the red line becomes tangent to the blue curve.

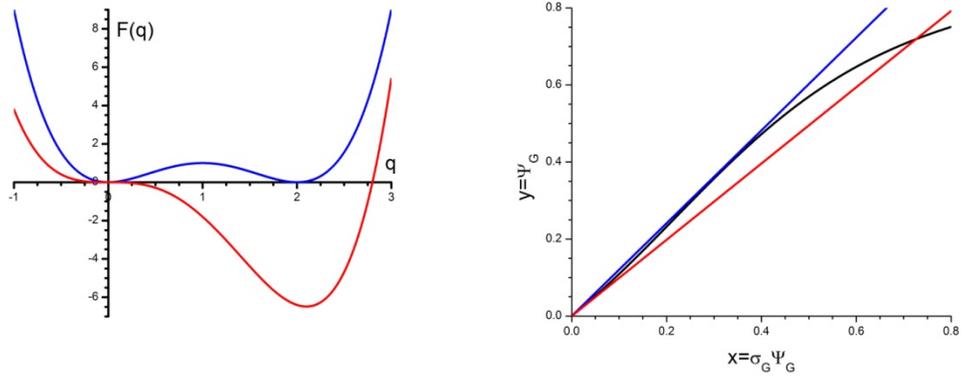

Figure 3. A comparison between (left panel) an illustration of the Landau theory of a first order transition and (right panel) an illustration of the bifurcation conditions Eq. (3.5) for the liquid to crystal phase transition and Eq. (3.7) for the limit of stability.

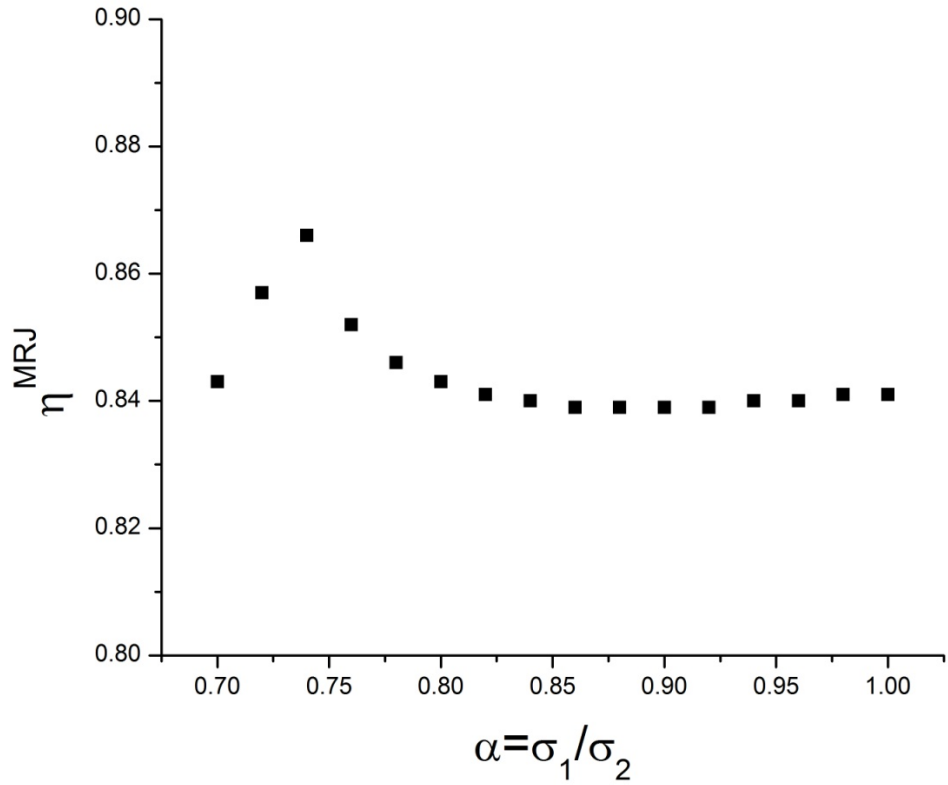

Figure 4. Calculated values of $\eta^{MRJ}$ for a binary hard disc mixture with composition $x_1 = x_2 = 0.5$, for the diameter ratio range $1 \geq \alpha \geq 0.7$. The calculations of the composition dependent structure factors of the binary mixture were based on the procedure proposed by Barrat et al [24] with Eq. (4.7).

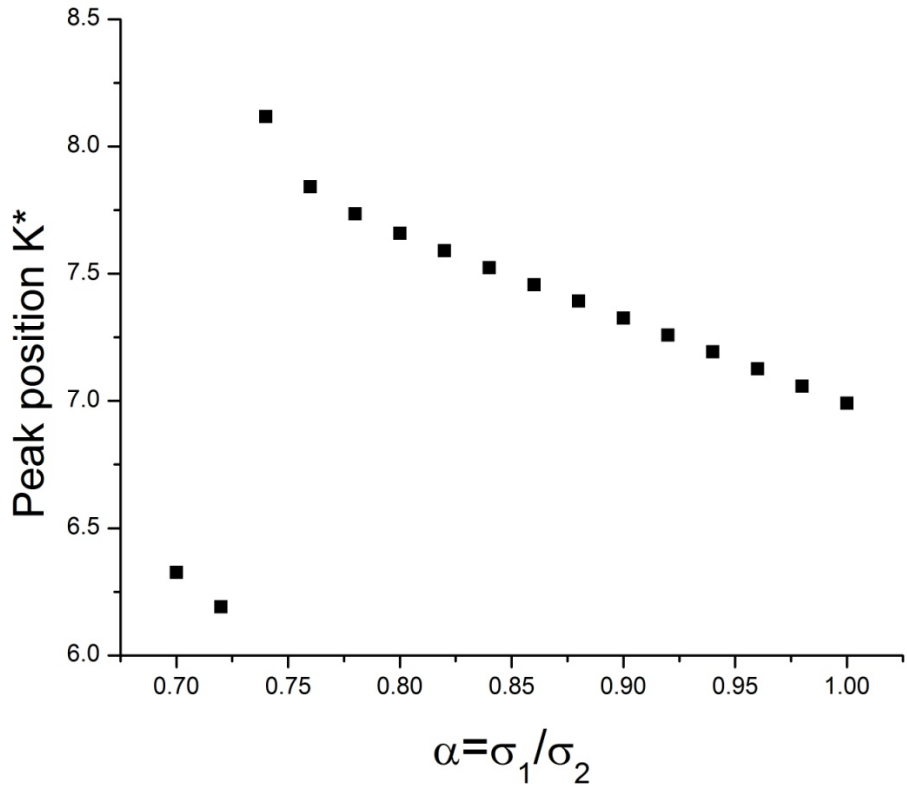

Figure 5. Calculated values of the peak position of $\lambda_s(k)$ for a binary hard disc mixture with composition $x_1 = x_2 = 0.5$, for the diameter ratio range $1 \geq \alpha \geq 0.7$. The calculations of the composition dependent structure factors of the binary mixture were based on the procedure proposed by Barrat et al [24] with Eq. (4.7).

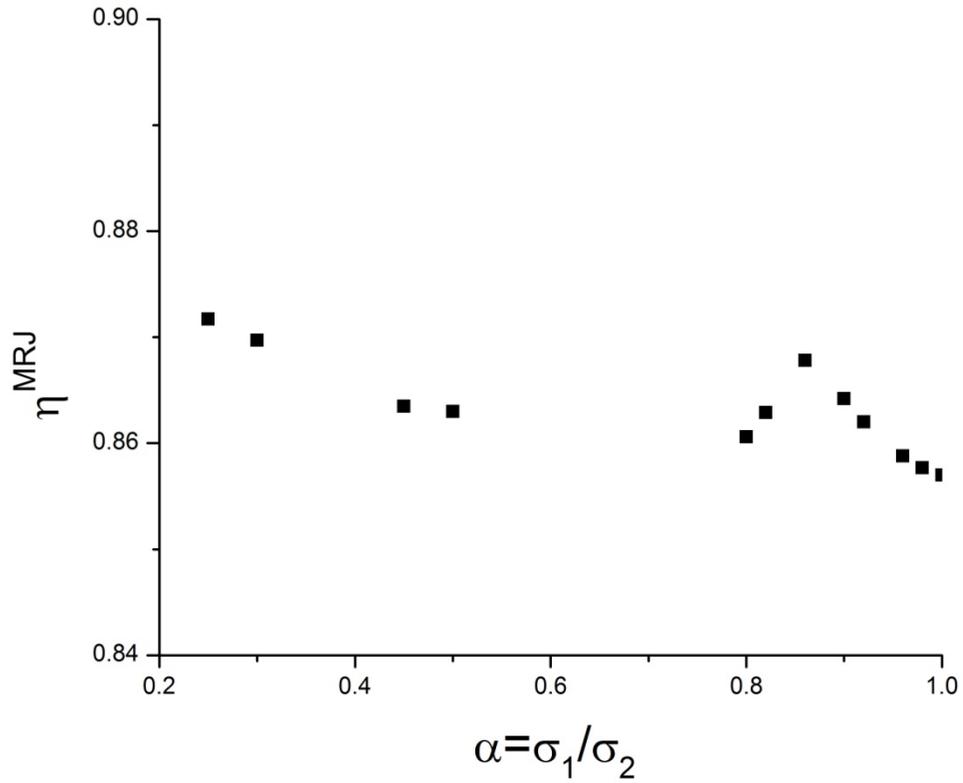

Figure 6. . Calculated values of $\eta^{MRJ}$ for a binary hard disc mixture with mole fraction composition given by the values of $\eta_{CP}$ determined by Fejes Toth for the densest covering of the plane. The calculations of the composition dependent structure factors of the binary mixture were based on the procedure proposed by Barrat et al [24] with Eq. (4.8).

# References

1. J. D. Bernal and J Mason, Nature **188**, 910 (1960)
2. G. Y. Onoda and E. G. Liniger, phys. Rev. Lett. **64**, 2727 (1990)
3. H. M. Jaeger, S. R. Nagel and R. P. Behringer, Rev. Mod. Phys. **68**, 1259 (1996)
4. A. J. Liu and S. R. Nagel, Nature **396**, 21 (1998)
5. G. Parisi and F. Zamponi, Rev. Mod. Phys. **82**, 789 (2010)
6. F. H. Stillinger, Science **267**, 1935 (1995)
7. A. Donev, F. H. Stillinger and S. Torquato, Phys. Rev. Lett. **96**, 225502 (2006)
8. M. Clusel, E. I. Corwin, A. O. N. Siemens and J. Brujic, Nature **460**, 611 (2009)
9. C. Song, P. Wang and H. A. Makse, Nature **453**, 629 (2008)
10. S. Torquato and F. H. Stillinger, Rev. Mod. Phys. **82**, 2633 (2010)
11. S. Torquato, T. M. Truskett and P. G. Debenedetti, Phys. Rev. Lett. **84**, 2064 (2000)
12. P. M. Chaikin and T. C. Lubensky, *Principles of condensed matter physics* (University Press, Cambridge, 1997), Chapter 4.
13. C. H. Mak, Phys. Rev. E **73**, 065104 (2006)
14. X. L. Xu and S. A. Rice, Phys. Rev. E **78**, 011602 (2008)
15. X. L. Xu and S. A. Rice, Proc. R. Soc. London, Ser. A **464** 65 (2008)
16. X. Cheng, Phys. Rev. E **81**, 031301 (2010)
17. D. Bideau, A. Gervois, L. Oger and J. P. Troadec, J. Physique **47**, 1697 (1986)
18. G. C. Barker and M. J. Grimson, J. Phys. Condens. Matter **1**, 2779 (1989)
19. E. A. Arinshtein, Dokl. Akad. Nauk SSSR **112**, 615 (1957)
20. F. H. Stillinger and F. P. Buff, J. Chem. Phys. **37**, 1 (1962)
21. J. E. Mayer, J. Chem. Phys. **15**, 187 (1947)
22. B. Bagchi, C. Cerjan and S. A. Rice, Phys. Rev. B **28**, 6411 (1983)
23. M. Baus and J. L. Colot, Phys. Rev. A **36**, 3912 (1987)
24. J-L. Barrat, H. Xu, J-P. Hansen and M. Baus, J. Phys. C: Solid State Phys. **21** 3165 (1988)
25. F. Saija, G. Fiumara and P. V. Giaquinta, Mol. Phys. **87**, 991 (1996)
26. R. J. Speedy, J. Chem. Phys. **110**, 4559 (1999)
27. J. T. Jenkins and F. Mancini, J. Appl. Phys. **54**, 27 (1987)

28. A. Santos, S. B. Yuste and M. Lopez de Haro, J. Chem. Phys. **117**, 5785 (2002)
29. D. Henderson, Mol. Phys. **30**, 971 (1975)
30. Y. Rosenfeld, Phys. Rev. **A 42**, 5978 (1990).
31. L. Fejes Toth, *Regular Figures*, MacMillan, New York (1960).
32. C. E. Zachary, Y. Jiao and S. Torquato, arXiv:1008.2548v1, 15 August 2010.
33. J. J. Kozak, J. Brzezinski and S. A. Rice, J. Phys. Chem. B **112**, 16059 (2008).
34. D. E. G. Williams, Phys. Rev. **E 57**, 7344 (1998).